\documentclass[pra,aps,preprint,showpacs,superscriptaddress]{revtex4}
\usepackage{dcolumn}
\usepackage{bm}
\usepackage{amsmath,amssymb}
\usepackage{graphicx}

\usepackage{color}

\begin{document}

\title{Theory of attosecond transient absorption spectroscopy of strong-field-generated ions}
\author{Robin Santra}
\affiliation{Center for Free-Electron Laser Science, DESY, Notkestra{\ss}e 85, 22607 Hamburg, Germany}
\affiliation{Department of Physics, University of Hamburg, Jungiusstra{\ss}e 9, 20355 Hamburg, Germany}
\affiliation{Kavli Institute for Theoretical Physics, University of California, Santa Barbara, CA 93106, USA}
\author{Vladislav S. Yakovlev}
\affiliation{Department f\"{u}r Physik, Ludwig-Maximilians-Universit\"{a}t, Am Coulombwall 1, D-85748 Garching, Germany}
\affiliation{Max-Planck-Institut f\"{u}r Quantenoptik, Hans-Kopfermann-Str. 1, D-85748 Garching, Germany}
\author{Thomas Pfeifer}
\affiliation{Max Planck Institute for Nuclear Physics, Saupfercheckweg 1, 69117 Heidelberg, Germany}
\author{Zhi-Heng Loh}
\affiliation{Departments of Chemistry and Physics, University of California, Berkeley, CA 94720, USA}
\affiliation{Chemical Sciences Division, Lawrence Berkeley National Laboratory, Berkeley, CA 94720, USA}
\date{\today}
\begin{abstract}
Strong-field ionization generally produces ions in a superposition of ionic eigenstates.  This superposition is
generally not fully coherent und must be described in terms of a density matrix.  A recent experiment 
[E. Goulielmakis {\em et al.}, Nature {\bf 466}, 739 (2010)] employed attosecond transient absorption spectroscopy 
to determine the density matrix of strong-field-generated Kr$^+$ ions.  The experimentally observed degree of
coherence of the strong-field-generated Kr$^+$ ions is well reproduced by a recently developed multichannel 
strong-field-ionization theory.  But there is significant disagreement between experiment and theory with respect 
to the degree of alignment of the Kr$^+$ ions.  In the present paper, the theory underlying attosecond transient 
absorption spectroscopy of strong-field-generated ions is developed.  The theory is formulated in such a way that the 
nonperturbative nature of the strong-field-ionization process is systematically taken into account.  The impact of 
attosecond pulse propagation effects on the interpretation of experimental data is investigated both analytically 
and numerically.  It is shown that attosecond pulse propagation effects cannot explain why the experimentally
determined degree of alignment of strong-field-generated Kr$^+$ ions is much smaller than predicted by existing theory.
\end{abstract}
\maketitle

%32.80.Rm 	Multiphoton ionization and excitation to highly excited states 
%82.53.Kp       Coherent spectroscopy of atoms and molecules in physical chemistry and chemical physics
%31.15.A- 	Ab initio calculations
%42.65.Re 	Ultrafast processes; optical pulse generation and pulse compression 

\section{Introduction}
\label{sec1}

The interaction of matter with a light pulse sets electrons in motion. The associated dynamics frequently occur 
on time scales comparable to or smaller than a femtosecond. The time-domain observation of such fast phenomena 
was impossible until % the advent of attosecond science
a few years ago, when the first pump-probe measurements with attosecond 
extreme-ultraviolet (EUV) pulses were demonstrated \cite{Drescher_Nature_2002}. Since then, several types of 
attosecond spectroscopy have been established or proposed theoretically \cite{Krausz_RMP_2009}.
% Most of these approaches take advantage of the fact that EUV pulses are produced by an intense near-infrared (NIR)
% laser pulse as a result of high harmonic generation, which inherently leads to a precise synchronization of the EUV
% and NIR fields.
Most of these approaches employ an intense near-infrared (NIR) laser pulse that either induces or probes
  dynamics of interest. Combined with an EUV pulse, an attosecond pump-probe measurement is performed by varying
the delay between the pulses and observing the results of their interaction with a system under study.  One of the
most prominent examples is attosecond streaking
\cite{Itatani_PRL_2002,Kienberger_Nature_2004,Eckle_NaturePhysics_2008}, where photoelectron spectra are measured
for a set of delays.
% Recently, attosecond streaking measurements have revealed that even such a fundamental phenomenon as single-photon
% ionization of an atom is far from being completely understood \cite{Schultze_Science_2010}.
%While attosecond streaking is
Attosecond streaking has proved to be a very powerful technique for measuring electron dynamics triggered by an
  attosecond pulse \cite{Schultze_Science_2010}, but it is less suited to probing strong-field dynamics with an
EUV pulse.  Historically, the first time-resolved measurements of electron dynamics induced by a strong light field
were performed with the aid of attosecond tunneling spectroscopy \cite{Uiberacker_Nature_2007}, where ions, rather
than electrons, were measured for different delays between the EUV and NIR pulses.  Very recently, another promising
technique---\emph{attosecond transient absorption spectroscopy}---has been established: by measuring the
transmission of an EUV pulse through a gas ionized by an NIR pulse, the motion of ionic valence-shell electrons has
been observed with attosecond resolution \cite{Goulielmakis_Nature_2010}.

The earliest implementation of transient absorption spectroscopy for the study of ultrafast dynamics involved 
detecting the change in absorbance of a narrowband probe pulse as a function of pump-probe time delay 
\cite{DaRo87,RoDa88,ScJo93,JoBr95}. In experiments that elucidate ultrafast molecular dynamics, tuning the 
central wavelength of the probe pulse tracks the motion of the nuclear wave packet into and out of different 
regions of the excited-state potential energy surface. Advances in laser pump-synchrotron probe techniques 
have allowed time-resolved x-ray absorption spectra to be collected in this fashion 
\cite{SaBr03,BrCh04,YoAr06,SoAr07,PeBu08,ChJa01,Chen05}. 

Mathies and Shank introduced a variant of transient absorption spectroscopy in which the photoexcitation of
a sample by an optical pump pulse is followed by probing with a spectrally broadband, few-cycle visible pulse
\cite{PoMa92,MaCr88,PoLe90,WaSc94}. The probe pulse that is transmitted through the sample is spectrally dispersed
by a spectrometer. In this method, the time resolution is mainly determined by the duration of the pump and
  probe pulses, and it is independent of the spectral resolution.
%dictated by the temporal cross-correlation between the pump and probe pulses,
%which is independent of the spectral resolution.
Hence, compared to earlier approaches that employed narrowband
probe pulses, spectrally
% dispersed
resolved transient absorption spectroscopy with broadband probe pulses offers the obvious
advantages of high time resolution and high spectral resolution. In the visible, state-of-the-art experiments on samples
of biological relevance have been done with sub-5-fs pump and probe pulses \cite{KoSa01}. This technique has also been
applied to femtosecond studies of novel materials \cite{GrCe98,Klim00}. For an up-to-date summary of the development of
the optical transient absorption technique, see Ref.~\cite{MePu09}. Recently, femtosecond transient absorption
spectroscopy was extended to shorter probe wavelengths by employing laser-produced bremsstrahlung x-rays \cite{RaWi96} and
EUV high-order harmonics \cite{LoKh07,SeSp07,LoLe08,LoGr08}, as well as femtosecond synchrotron x-rays \cite{CaRi05}.

Transient absorption spectroscopy provided direct evidence for hole alignment in strong-field-generated atomic ions
\cite{YoAr06}.  The ion alignment dynamics driven by electron-ion collisions in a strong-field-generated plasma were
studied in Ref.~\cite{HoPe07}.  However, it was not clear at the time whether strong-field ionization leads to the
formation of a coherent superposition of ionic eigenstates (in cases where not only the ionic ground state is
populated).  In other words, it was not clear whether strong-field-generated ions undergo any coherent intraatomic
dynamics.  Theory indicated that it is generally not possible to describe strong-field-generated ions in terms of a
perfectly coherent wave packet; a description in terms of a density matrix is required \cite{SaDu06,RoSa09}.  Using
transient absorption spectroscopy in combination with theory, the diagonal elements of the density matrix of
strong-field-generated atomic ions were measured \cite{LoKh07,SoAr07}.  The diagonal elements of the density matrix
(represented in the ionic eigenbasis) are the populations of the various ionic eigenstates, generated by a strong NIR
field. The measurement of the entire ion density matrix---including the coherences, i.e., the off-diagonal elements of the
ion density matrix---was made possible by attosecond transient absorption spectroscopy \cite{Goulielmakis_Nature_2010}.

In the present paper, we develop a theoretical description of attosecond transient absorption spectroscopy of 
strong-field-generated ions.  Our approach allows us to treat the pump step, i.e., the interaction with a strong NIR 
field, in a completely nonperturbative fashion.  We assume throughout that the attosecond EUV probe pulse has no temporal 
overlap with the strong NIR pump pulse.  In Sec.~\ref{sec2}, we derive the basic expressions underlying the theory.  
The analysis of the experiment of Ref.~\cite{Goulielmakis_Nature_2010} becomes particularly transparent if it is assumed
that the attosecond probe pulse remains sufficiently short as it propagates through the NIR-modified target medium.  
This short-pulse approximation is discussed in Sec.~\ref{sec3}.  Numerical EUV pulse propagation calculations in 
Sec.~\ref{sec4} allow us to assess the validity of the short-pulse approximation under the conditions of 
Ref.~\cite{Goulielmakis_Nature_2010}.  The distortion of the NIR pump pulse by the target medium is not considered.
Section~\ref{sec5} concludes. Atomic units are used throughout, unless otherwise noted.

\section{General considerations}
\label{sec2}

\subsection{Atomic response}
\label{sec2a}

We consider a semiclassical description of the interaction of the pump and probe pulses with the atoms in the target gas.
Both pulses are approximated by transverse, infinite plane waves propagating along the $x$ axis, and are assumed to be
linearly polarized along the $z$ axis.  As mentioned in the introduction, the NIR pulse is assumed to remain unmodified as
it propagates through the gas.  Thus, neglecting diffraction, the NIR electric field may be written as
$\mathcal{E}_{\mathrm{NIR}}(t_L - x/c)$, where $t_L$ is the time measured in the laboratory frame and 
$c$ is the vacuum speed of light.  For an atom at
position $x$, it is convenient to introduce the local time $t = t_L - x/c$.  Hence, in the electric dipole approximation,
the Hamiltonian for an atom at position $x$ reads
\begin{equation}
\label{eq1}
\hat{H}(t) = \hat{H}_0 - E_0 - \mathcal{E}_{\mathrm{NIR}}(t)  \hat{Z} - \mathcal{E}_{\mathrm{EUV}}(x,t+x/c) \hat{Z}.
\end{equation}
Here, $\hat{H}_0$ is the unperturbed atomic Hamiltonian, $E_0$ is the atomic ground-state energy, $\hat{Z}$ is the
$z$ component of the electric dipole operator, and $\mathcal{E}_{\mathrm{EUV}}(x,t_L)$ is the EUV electric field.  In
order to calculate the EUV-induced polarization response of the atoms, we need to solve the time-dependent Schr\"{o}dinger
equation
\begin{equation}
\label{eq2}
i \frac{\partial}{\partial t} |\Psi,t\rangle = \hat{H}(t) |\Psi,t\rangle.
\end{equation}

Let us assume we have solved the $x$-independent, NIR-only problem
\begin{equation}
\label{eq3}
i \frac{\partial}{\partial t} \hat{U}_{\mathrm{NIR}}(t,-\infty)= \left\{\hat{H}_0 - E_0 - \mathcal{E}_{\mathrm{NIR}}(t)  \hat{Z}\right\}\hat{U}_{\mathrm{NIR}}(t,-\infty).
\end{equation}
A suitable initial condition for the time evolution operator is
\begin{equation}
\label{eq4}
\hat{U}_{\mathrm{NIR}}(t,-\infty) \rightarrow \exp{\left\{-i(\hat{H}_0 - E_0)t\right\}}
\mbox{\ as}\ 
 t \rightarrow -\infty.
\end{equation}
Therefore, in the absence of the probe pulse, a solution to Eq.~\eqref{eq2} is
\begin{equation}
\label{eq4a}
|\Psi_{\mathrm{NIR}},t\rangle \equiv \hat{U}_{\mathrm{NIR}}(t,-\infty)|\Psi_0\rangle.
\end{equation}
Here, $|\Psi_0\rangle$ is the initial state of the atom, assumed to be the ground state.  

In order to take into consideration the effect of the probe pulse, we make the ansatz
\begin{equation}
\label{eq5}
|\Psi,t\rangle = |\Psi_{\mathrm{NIR}},t\rangle + |\Psi,t\rangle^{(1)} + \ldots,
\end{equation}
where $|\Psi,t\rangle^{(1)}$ is a correction that is of first order with respect to $\mathcal{E}_{\mathrm{EUV}}(x,t+x/c)$.
It follows that
\begin{equation}
\label{eq6}
i \frac{\partial}{\partial t} |\Psi,t\rangle^{(1)} = \left\{\hat{H}_0 - E_0 - \mathcal{E}_{\mathrm{NIR}}(t)
  \hat{Z}\right\} |\Psi,t\rangle^{(1)}
-\mathcal{E}_{\mathrm{EUV}}(x,t+x/c) \hat{Z} |\Psi_{\mathrm{NIR}},t\rangle.
\end{equation}
This equation can be integrated analytically.  The result reads
\begin{equation}
\label{eq7}
|\Psi,t\rangle^{(1)} = i \int_{-\infty}^{t} dt' \hat{U}_{\mathrm{NIR}}(t,t')\hat{Z}\mathcal{E}_{\mathrm{EUV}}(x,t'+x/c)|\Psi_{\mathrm{NIR}},t'\rangle,
\end{equation}
%The time evolution operator $\hat{U}_{\mathrm{NIR}}(t,t')$ is simply
%$\hat{U}_{\mathrm{NIR}}(t,-\infty)\hat{U}_{\mathrm{NIR}}^{\dag}(t',-\infty)$.
where $\hat{U}_{\mathrm{NIR}}(t,t')=\hat{U}_{\mathrm{NIR}}(t,-\infty)\hat{U}_{\mathrm{NIR}}^{\dag}(t',-\infty)$.

Let us now calculate the polarization along the $z$ axis:
\begin{eqnarray}
\label{eq8}
P(x,t+x/c) & = & n_{\mathrm{AT}}\langle\Psi,t|\hat{Z}|\Psi,t\rangle \\
     & = & P_{\mathrm{HG}}(x,t+x/c) + P^{(1)}(x,t+x/c) + \dots. \nonumber
\end{eqnarray}
In this expression, $n_{\mathrm{AT}}$ is the atomic number density, 
\begin{equation}
\label{eq9}
P_{\mathrm{HG}}(x,t+x/c) = n_{\mathrm{AT}} \langle\Psi_{\mathrm{NIR}},t|\hat{Z}|\Psi_{\mathrm{NIR}},t\rangle 
\end{equation}
describes harmonic generation driven by the NIR pulse (no EUV pulse present), and 
\begin{equation}
\label{eq10}
P^{(1)}(x,t+x/c) = n_{\mathrm{AT}} \langle\Psi_{\mathrm{NIR}},t|\hat{Z}|\Psi,t\rangle^{(1)} + \mathrm{c.c.}
\end{equation}
is the polarization correction to first order with respect to $\mathcal{E}_{\mathrm{EUV}}(x,t+x/c)$.

Using Eq.~\eqref{eq7}, the first-order polarization correction may be written as
\begin{multline}
\label{eq11}
P^{(1)}(x,t+x/c) = i n_{\mathrm{AT}} \int_{-\infty}^{t} dt' \biggl\{
\mathcal{E}_{\mathrm{EUV}}(x,t'+x/c) \\
\times \langle\Psi_{\mathrm{NIR}},t|\hat{Z}\hat{U}_{\mathrm{NIR}}(t,t')\hat{Z}
|\Psi_{\mathrm{NIR}},t'\rangle \biggr\} + \mathrm{c.c.}
\end{multline}
This result, which is valid for arbitrary pump intensities, generalizes Eq.~(17b) in Ref.~\cite{PoLe90} and shows 
that Eq.~(21) in that paper is not quite correct \footnote{Equation~(21) in Ref.~\cite{PoLe90} should read
\[\sigma(\omega,t_0) = \frac{4\pi\omega}{\hbar c}{\mathrm{Re}}\biggl[\int_0^{\infty}dt \; e^{-\gamma_{21}t}
e^{i\omega t} \times \langle i(t_0)| e^{i h_1 t/\hbar} \mu_{21}^{\ast} e^{-i h_2 t/\hbar} \mu_{21} |i(t_0)\rangle\biggr].\]}.  
Note that the integrand in Eq.~\eqref{eq11} depends explicitly 
on $t$ via $\langle\Psi_{\mathrm{NIR}},t|$ and $\hat{U}_{\mathrm{NIR}}(t,t')$.  
Therefore, even if the EUV pulse may, effectively, be approximated
by a delta function centered at, say, $t' = t_{\mathrm{EUV}}$, the polarization induced by the EUV pulse contains information not only
on atomic properties at the instant of the EUV pulse, but, in principle, also on atomic properties after the EUV pulse (for $t > t_{\mathrm{EUV}}$).
As we will see, this causes no difficulty for the situation considered in this paper.

We assume that the EUV probe pulse comes after the NIR pump pulse, so that the NIR pulse does not affect the electronic states reached via 
EUV photoabsorption.  Hence, we have, for the time evolution operator in Eq.~\eqref{eq11}, 
\begin{equation}
\label{eq13}
\hat{U}_{\mathrm{NIR}}(t,t') = e^{-i(\hat{H}_0 - E_0)(t-t')}.
\end{equation}
Let $|I\rangle$ denote an eigenstate of the unperturbed atomic Hamiltonian $\hat{H}_0$ with eigenenergy $E_I$.  We may then expand the 
NIR-only state vector, after the NIR pulse, as follows:
\begin{equation}
\label{eq12}
|\Psi_{\mathrm{NIR}},t\rangle = \sum_I \alpha_I e^{-i(E_I - E_0)t}|I\rangle.
\end{equation}
The expansion coefficients $\alpha_I$ are time-independent.  Thus, Eq.~\eqref{eq11} goes over into
\begin{multline}
\label{eq13a}
P^{(1)}(x,t+x/c) = i n_{\mathrm{AT}} \sum_{I,I'} \alpha_I^{\ast} \alpha_{I'}
\sum_F \langle I|\hat{Z}|F\rangle \langle F|\hat{Z}|I'\rangle \\
 \times \int_{-\infty}^{t} dt' \mathcal{E}_{\mathrm{EUV}}(x,t'+x/c) e^{-i(E_F - E_I)(t-t') + i(E_I - E_{I'})t'}
+ \mathrm{c.c.},
\end{multline}
where $|F\rangle$ is an eigenstate of $\hat{H}_0$ with eigenenergy $E_F$, which is, in general, assumed to be complex.  More precisely,
the imaginary part of $E_F$ is either zero or negative.

\subsection{EUV pulse propagation}
\label{sec2b}

We assume that the high harmonics generated by the NIR pulse [Eq.~\eqref{eq9}] do not overlap with the spectral range of
the EUV pulse. 
%Hence, after neglecting the impact of electric currents on the magnetic field, the following equation of
%motion for the EUV electric field $\mathcal{E}_{\mathrm{EUV}}(x,t_L)$ may be derived from Maxwell's equations:
Hence, the propagation of the EUV electric field through the medium can be described by the following
scalar wave equation:
\begin{equation}
\label{eq13b}
\left(\frac{\partial^2}{\partial x^2} - \frac{1}{c^2}\frac{\partial^2}{\partial t_L^2}\right)\mathcal{E}_{\mathrm{EUV}}(x,t_L)
= \frac{4\pi}{c^2}\frac{\partial^2}{\partial t_L^2} P^{(1)}(x,t_L).
\end{equation}
With the Fourier representations
\begin{equation}
\label{eq13c}
\mathcal{E}_{\mathrm{EUV}}(x,t_L) = \int_0^{\infty} \frac{d\omega}{2\pi} 
\left\{\tilde{\mathcal{E}}_{\mathrm{EUV}}(x,\omega) e^{-i\omega (t_L - x/c)} + \mathrm{c.c.}\right\},
\end{equation}
\begin{equation}
\label{eq13d}
P^{(1)}(x,t_L) = \int_0^{\infty} \frac{d\omega}{2\pi}
\left\{\tilde{P}^{(1)}(x,\omega) e^{-i\omega (t_L - x/c)} + \mathrm{c.c.}\right\},
\end{equation}
it follows from Eq.~\eqref{eq13b} that
\begin{equation}
\label{eq13e}
\left(\frac{\partial^2}{\partial x^2} + 2 i \frac{\omega}{c}\frac{\partial}{\partial x}\right)\tilde{\mathcal{E}}_{\mathrm{EUV}}(x,\omega)
= - 4\pi \frac{\omega^2}{c^2}\tilde{P}^{(1)}(x,\omega).
\end{equation}
In the situation considered, the spatial derivative of the electric field amplitude at a given $\omega$ changes slowly 
over a wavelength $2\pi c/\omega$.  This allows us to neglect in Eq.~(\ref{eq13e}) the second derivative with respect to $x$.
Thus, the differential equation governing the spatial evolution of $\tilde{\mathcal{E}}_{\mathrm{EUV}}(x,\omega)$ reads
\begin{equation}
\label{eq13f}
\frac{\partial}{\partial x}\tilde{\mathcal{E}}_{\mathrm{EUV}}(x,\omega) = 2\pi i \frac{\omega}{c}\tilde{P}^{(1)}(x,\omega).
\end{equation}

Equation (\ref{eq13f}) can be
% immediately integrated
integrated analytically if $\tilde{P}^{(1)}(x,\omega)$ is proportional to
$\tilde{\mathcal{E}}_{\mathrm{EUV}}(x,\omega)$:
\begin{equation}
\label{eq13g}
\tilde{P}^{(1)}(x,\omega) = \chi^{(1)}(x,\omega) \tilde{\mathcal{E}}_{\mathrm{EUV}}(x,\omega).
\end{equation}
Here, $\chi^{(1)}(x,\omega)$ is the linear susceptibility.  If Eq.~\eqref{eq13g} is valid, it follows from Eq.~\eqref{eq13f} that
\begin{equation}
\label{eq13h}
\tilde{\mathcal{E}}_{\mathrm{EUV}}(x,\omega) = \tilde{\mathcal{E}}_{\mathrm{EUV}}(x_0,\omega) 
\exp{\left\{2\pi i \frac{\omega}{c} \int_{x_0}^x dx' \chi^{(1)}(x',\omega)\right\}}.
\end{equation}
In the attosecond transient absorption experiment described in Ref.~\cite{Goulielmakis_Nature_2010}, 
the EUV radiation transmitted through the sample (length $L$) was spectrally 
dispersed and analyzed.  The detected signal is then proportional to
\begin{equation}
\label{eq13i}
|\tilde{\mathcal{E}}_{\mathrm{EUV}}(L,\omega)|^2 = |\tilde{\mathcal{E}}_{\mathrm{EUV}}(0,\omega)|^2
e^{-4\pi \frac{\omega}{c} \int_0^L dx \mathrm{Im}[\chi^{(1)}(x,\omega)]},
\end{equation}
which, for a homogeneous target medium, goes over into
\begin{equation}
\label{eq13j}
|\tilde{\mathcal{E}}_{\mathrm{EUV}}(L,\omega)|^2 = |\tilde{\mathcal{E}}_{\mathrm{EUV}}(0,\omega)|^2
e^{-4\pi \frac{\omega}{c} L \mathrm{Im}[\chi^{(1)}(\omega)]}.
\end{equation}
Equation (\ref{eq13j}) is Beer's law.

To understand what determines the validity of Eq.~\eqref{eq13g}, and thus the applicability of Beer's law, 
we calculate, using Eqs. (\ref{eq13a}), (\ref{eq13c}), and (\ref{eq13d}), the Fourier transform of the 
EUV-induced polarization:
% for $\omega>0$:
\begin{multline}
\label{eq13k}
\tilde{P}^{(1)}(x,\omega) = \int_{-\infty}^{\infty} dt P^{(1)}(x,t+x/c)e^{i\omega t} \\
= n_{\mathrm{AT}} \sum_{I,I'} \alpha_I^{\ast} \alpha_{I'} 
\sum_F \langle I|\hat{Z}|F\rangle \langle F|\hat{Z}|I'\rangle
\biggl\{\frac{1}{E_F - E_I - \omega} \\+
\frac{1}{E_F^{\ast} - E_{I'} + \omega}\biggr\}
\tilde{\mathcal{E}}_{\mathrm{EUV}}(x,\omega+E_I-E_{I'}).
\end{multline}
Therefore, the ratio $\tilde{P}^{(1)}(x,\omega)/\tilde{\mathcal{E}}_{\mathrm{EUV}}(x,\omega)$ is independent of the EUV electric field---i.e., 
a well-defined linear susceptibility is obtained---only if 
\begin{equation}
\label{eq13l}
\frac{\tilde{\mathcal{E}}_{\mathrm{EUV}}(x,\omega+E_I-E_{I'})}{\tilde{\mathcal{E}}_{\mathrm{EUV}}(x,\omega)} =
 \mathrm{const.}
\end{equation}
This condition is always satisfied if only terms with $I'=I$ contribute to the polarization response in
  Eq.~\eqref{eq13k}. Generally, this is not the case. Still, Eq.~\eqref{eq13l} serves as a good approximation
%If in Eq.~\eqref{eq13k} for the polarization only terms with $I'=I$ play a role, then Eq.~\eqref{eq13l} is satisfied.
%Generally, however, if off-diagonal terms cannot be neglected, Eq.~\eqref{eq13l} can only be satisfied,
% independent of $x$ and $\omega$,
if the EUV pulse is much shorter than $2\pi/(E_I-E_{I'})$ for $I'\ne I$.  Furthermore, for Beer's law to be rigorously
  applicable, this short pulse must not undergo any significant distortion as it propagates through the dense gas, so
that the EUV pulse remains short in comparison to the dynamical time scales characterizing the target medium prepared by
the NIR pump pulse. If Eq.~\eqref{eq13l} is not satisfied, the propagation equation \eqref{eq13f} must be solved
  numerically. In the following sections, we discuss both regimes in more detail.

% In the following section, we ensure the validity of Eq.~\eqref{eq13l}---and thus of Eqs. (\ref{eq13i}) and
% (\ref{eq13j})---by assuming that the EUV pulse may be represented by a delta function.  Numerical EUV pulse propagation
% calculations for a finite EUV pulse duration are described in Sec.~\ref{sec4}.

\subsection{The configuration expansion}
\label{sec2c}
So far, our treatment of the electronic-structure problem has been general.  In order to describe resonant EUV absorption
by laser-generated ions, we now adopt the mean-field model discussed in Ref.~\cite{RoSa09} and write the NIR-only
state vector after the NIR pulse in terms of Slater determinants:
\begin{eqnarray}
\label{eq16}
|\Psi_{\mathrm{NIR}},t\rangle & = & \sum_I \alpha_I e^{-i(E_I - E_0)t}|I\rangle \\
& = & \alpha_0 |\Phi_0\rangle + \sum_i\sum_a \alpha_i^a e^{-i(\varepsilon_a-\varepsilon_i)t}|\Phi_i^a\rangle. \nonumber
\end{eqnarray}
In this approximation, the set $\{|I\rangle\}$ contains the ground-state determinant $|\Phi_0\rangle$ and all
particle-hole configurations $|\Phi_i^a\rangle$ obtained from $|\Phi_0\rangle$ by exciting or ionizing an
electron from an occupied spin orbital $i$ (a ``hole'' orbital) to an unoccupied spin orbital $a$ (a ``particle'' orbital)
\cite{FeWa71,Matt92,MaYo95}. A particle-hole configuration corresponds to an excited or ionized electron plus 
an ion core with a hole in some shell that is fully occupied in the ground state of the neutral atom.
The orbital energies of the hole and the particle are denoted by $\varepsilon_i$ and $\varepsilon_a$, respectively.
% ($\varepsilon_a$ is the orbital energy of the particle, $\varepsilon_i$ is the orbital energy of the hole).
The numerical calculation of the coefficients $\alpha_0$ and $\alpha_i^a$ in Eq.~\eqref{eq16} is described in
Ref.~\cite{RoSa09}.
%Correspondingly,
Since $\hat{Z}$ is a one-body operator, the sum over the states $|F\rangle$ in Eqs.~\eqref{eq13a} and \eqref{eq13k}
extends over $|\Phi_0\rangle$, the particle-hole configurations, and the two-particle--two-hole configurations.  The
various terms that arise in this way describe processes such as EUV absorption by the neutral ground-state atoms and EUV
absorption by the ion core.

\begin{figure}[ht]
  \begin{center}
    \includegraphics[clip,width=0.8\hsize]{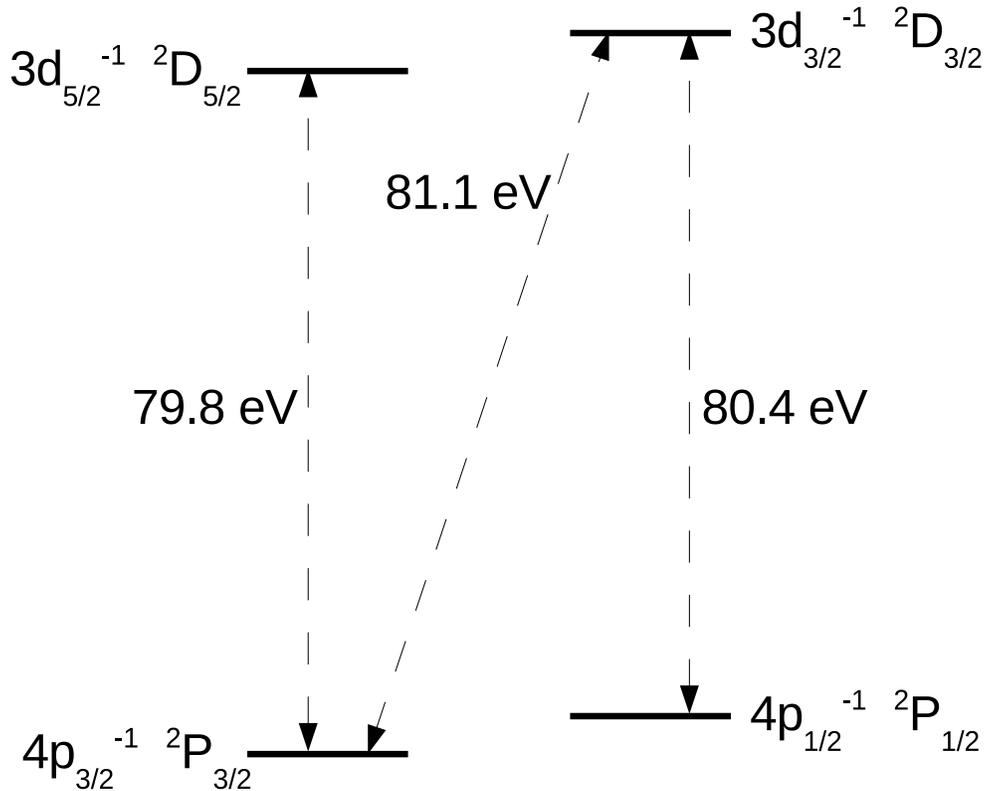}
    \caption{Atomic levels of Kr$^+$ associated with resonant absorption at photon energies near $80$~eV.  
    The notation $nl_j^{-1}$ indicates that, relative to the closed-shell ground state of the neutral atom, an electron 
    is missing (a hole is present) in the $nl_j$ subshell.}
    \label{fig1}
  \end{center}
\end{figure}

In the following, we focus on the experiment of Ref.~\cite{Goulielmakis_Nature_2010} (see Sec.~\ref{sec1}),
which studied the
%The most prominent
structures in the EUV absorption spectrum of strong-field-generated Kr$^+$ ions associated with 
exciting an inner-shell $3d$ electron in Kr$^+$ into the outer-valence $4p$ vacancy created by the NIR pulse.  
Resonant $3d-4p$ photoabsorption by the Kr$^+$ ion core was used to study the properties of the hole generated 
via strong-field ionization.  Since such transitions involve only hole orbitals, 
%leaves the particle unaffected; since only the so-called hole orbitals are involved, 
we refer to them
%these resonant transitions
as hole-hole transitions.  The relevant transitions are indicated in Fig.~\ref{fig1}.  The transition energies shown in
Fig.~\ref{fig1} were calculated using the multiconfiguration Dirac-Fock program package GRASP \cite{PaFr96}.  The
configurations that were included in the calculation are $4p_{3/2}^{-1}$, $4p_{1/2}^{-1}$, $4s_{1/2}^{-1}$,
$3d_{5/2}^{-1}$, and $3d_{3/2}^{-1}$.

We neglect that the EUV field might induce particle-particle transitions. This is an excellent approximation,
because the NIR-generated photoelectron interacts only weakly with the EUV field.  Since the electron excited or
ionized by the NIR pulse is a spectator in the hole-hole transitions of the ion core, one can describe the ions with
a reduced density matrix \cite{RoSa09}:
\begin{equation}
\label{eq17a}
\rho^{(\mathrm{ion})}_{ii'}(t) = e^{i(\varepsilon_i-\varepsilon_{i'})t} \sum_{a} \alpha_i^{a} {\alpha_{i'}^a}^{\ast}
= e^{i(\varepsilon_i-\varepsilon_{i'})t} \tilde{\rho}^{(\mathrm{ion})}_{ii'},
\end{equation}
where the summation is performed over all unoccupied orbitals. Since we consider the regime where the NIR and EUV
pulses do not overlap, the auxiliary matrix $\tilde{\rho}^{(\mathrm{ion})}_{ii'}$ is time-independent.

The polarization response of the ions to the field of the EUV pulse can be expressed via the density matrix by
  inserting the ansatz \eqref{eq16} for $|\Psi_{\mathrm{NIR}},t\rangle$ into Eq.~\eqref{eq11} and following the same steps in
  the derivation that led to Eq.~\eqref{eq13k}. This yields
\begin{multline}
  \label{eq:P1}
  \tilde{P}^{(1)}(x,\omega) =
  n_\mathrm{AT} \sum_{i,i'} \tilde{\rho}_{i i'}^{\mathrm{(ion)}}
  \sum_{\tilde{i}} z_{\tilde{i} i'} z_{i \tilde{i}}
  \biggl\{
    \frac{1}{\varepsilon_{i'} - \varepsilon_{\tilde{i}} - \omega} \\+
    \frac{1}{\varepsilon_{i} - \varepsilon_{\tilde{i}}^\ast + \omega}
  \biggr\}
  \tilde{\mathcal{E}}_\mathrm{EUV}(x, \omega+\varepsilon_i-\varepsilon_{i'}).
\end{multline}
Here, the absorption of an EUV photon fills a hole in orbital $i$ with an electron from orbital $\tilde{i}$, which has the
orbital energy $\varepsilon_{\tilde{i}}$. This process is described by the dipole matrix element $z_{i \tilde{i}}$. The
second dipole matrix element, $z_{\tilde{i} i'}$, describes the process upon which the hole created in orbital $\tilde{i}$
is filled by an electron from orbital $i'$.

In the following, we express the hole orbital energies in terms of ionization potentials, $I_i=-\varepsilon_i$ being defined as the
  minimum energy required to create an ion with a hole in orbital $i$ from a neutral atom in its ground
  state. Furthermore, since $\tilde{P}^{(1)}(x,-\omega)=\left(\tilde{P}^{(1)}(x,\omega)\right)^\ast$, it is sufficient to
  calculate the polarization response only for positive frequencies. Dropping the counter-rotating term in
  Eq.~\eqref{eq:P1} and using ionization potentials instead of orbital energies, we obtain
\begin{equation}
  \label{eq:P2}
  \tilde{P}^{(1)}(x,\omega>0) =
  n_\mathrm{AT} \sum_{i,i'} \tilde{\rho}_{i i'}^{\mathrm{(ion)}}
  \sum_{\tilde{i}}
    \frac{z_{\tilde{i} i'} z_{i \tilde{i}}}{I_{\tilde{i}} - i\frac{\Gamma_{\tilde{i}}}{2} - I_{i'} - \omega}
    \tilde{\mathcal{E}}_\mathrm{EUV}(x, \omega-I_i+I_{i'}),
\end{equation}
where $\Gamma_{\tilde{i}}$ is the decay width of the one-hole channel $\tilde{i}$. We emphasize that Eq.~\eqref{eq:P2}
is valid only if the EUV pulse comes after the NIR pulse.  Recall also that the NIR pulse is assumed to remain unmodified 
as it propagates through the gas.  If we dropped this assumption, then the ion density matrix would depend explicitly on the 
atomic position $x$ along the NIR pulse propagation axis.  The computationally expensive electronic wave-packet problem 
(\ref{eq3}) would then have to be solved not only once, but for every grid point used in the discretization of $x$.

%perform the summation over the particle states when evaluating Eq.~\eqref{eq15}.  Thus, after neglecting the
%counter-rotating term in Eq.~\eqref{eq15} and concentrating on the hole-hole transitions, the linear susceptibility may be
%written as

\section{Short-pulse approximation}
\label{sec3}

We now analyze the EUV-induced polarization assuming that the probe field may be approximated by a delta function centered at $t=t_{\mathrm{EUV}}$, i.e.,
\begin{equation}
\label{eq14}
\mathcal{E}_{\mathrm{EUV}}(x,t+x/c) \propto \delta(t-t_{\mathrm{EUV}}).
\end{equation}
Within this approximation, the Fourier transform of the EUV electric field is
\begin{equation}
\label{eq14a}
\tilde{\mathcal{E}}_{\mathrm{EUV}}(x,\omega) \propto e^{i\omega t_{\mathrm{EUV}}}.
\end{equation}
Therefore, we obtain from Eqs.~\eqref{eq13g}, \eqref{eq17a}, and \eqref{eq:P2} the linear susceptibility as 
\begin{eqnarray}
\label{eq17}
\chi^{(1)}(\omega>0) & = & \frac{\tilde{P}^{(1)}(x,\omega)}{\tilde{\mathcal{E}}_{\mathrm{EUV}}(x,\omega)} \nonumber \\
& = & n_{\mathrm{AT}} \sum_{i,i'} \rho^{(\mathrm{ion})}_{ii'}(t_{\mathrm{EUV}}) 
\sum_{\tilde{i}} \frac{z_{\tilde{i}i'}z_{i\tilde{i}}}{I_{\tilde{i}} - i \frac{\Gamma_{\tilde{i}}}{2} - I_{i'} - \omega}.
\end{eqnarray}

This equation shows that the linear susceptibility has poles that are simply related to transition energies of the ion core.  
The existence of an $x$-independent linear susceptibility allows one to use Beer's law [Eq.~\eqref{eq13j}] to calculate 
the spectrum of the EUV radiation transmitted through the target medium.  From a practical perspective, this means the following:  Under the 
conditions assumed, one obtains, by taking the logarithm of 
$|\tilde{\mathcal{E}}_{\mathrm{EUV}}(L,\omega)|^2/|\tilde{\mathcal{E}}_{\mathrm{EUV}}(0,\omega)|^2$, 
a quantity that is proportional to the EUV one-photon cross section
\begin{equation}
\label{eq18}
\sigma^{(1)}(\omega) = 4\pi \frac{\omega}{c}\frac{\mathrm{Im}[\chi^{(1)}(\omega)]}{n_{\mathrm{AT}}}.
\end{equation}

Now we are ready to give an explicit expression for the EUV one-photon cross section of strong-field-generated Kr${}^+$
  ions. Let $j$ be the total angular-momentum quantum number of an orbital hole created by the NIR pulse, and let $m$ be
the corresponding projection quantum number.  We exploit the fact that the reduced ion density matrix is diagonal in
$m$ \cite{RoSa09} and denote the ion density matrix elements by $\rho^{(m)}_{j,j'}$. Furthermore, it can be shown that
  terms containing $\rho^{(-m)}_{j,j'}$ give the same contributions to the polarization response as those containing
  $\rho^{(m)}_{j,j'}$.
% After the NIR pulse, the quantum-state populations $\rho^{(m)}_{j,j}$ are constant.
% Further, $\rho^{(-m)}_{j,j} = \rho^{(m)}_{j,j}$.  All nonzero off-diagonal
% elements are time-dependent and are trivially related \cite{RoSa09}.  It is therefore sufficient to select one of them.
We refer to the off-diagonal element $\rho^{(1/2)}_{3/2,1/2}(t_{\mathrm{EUV}})$ of the density matrix between the
  $4p_{3/2}^{-1}$, $m=+1/2$ and the $4p_{1/2}^{-1}$, $m=+1/2$ ionization channels as the \emph{coherence}. This element is
  equal to $\tilde{\rho}^{(1/2)}_{3/2,1/2} e^{-i(I_{4p_{3/2}} - I_{4p_{1/2}})t_{\mathrm{EUV}}}$ [cf. Eq.~\eqref{eq17a}],
i.e., $|\rho^{(1/2)}_{3/2,1/2}(t_{\mathrm{EUV}})| = |\tilde{\rho}^{(1/2)}_{3/2,1/2}| = \mathrm{const}$.  The complex
constant $\tilde{\rho}^{(1/2)}_{3/2,1/2}$ generally differs from zero, unless the statistical mixture described by the ion
density matrix is completely incoherent.  For a perfectly coherent hole wave packet, $|\tilde{\rho}^{(1/2)}_{3/2,1/2}|$
would equal $\sqrt{\rho^{(1/2)}_{3/2,3/2}\rho^{(1/2)}_{1/2,1/2}}$.  As demonstrated theoretically in Ref.~\cite{RoSa09}
and experimentally in Ref.~\cite{Goulielmakis_Nature_2010}, strong-field ionization does not in general produce perfectly
coherent hole wave packets.  In other words, generally $|\tilde{\rho}^{(1/2)}_{3/2,1/2}| <
\sqrt{\rho^{(1/2)}_{3/2,3/2}\rho^{(1/2)}_{1/2,1/2}}$.

Using Eq.~\eqref{eq18} and the notation just introduced, the EUV one-photon cross section associated with the hole-hole 
transitions reads, in the case of krypton,
\begin{multline}
\label{eq19}
\sigma^{(1)}(\omega) = 4\pi \frac{\omega}{c} \mathrm{Im}\Biggl\{
\frac{|\langle 4p_{3/2}^{-1}||D||3d_{5/2}^{-1}\rangle|^2}{I_{3d_{5/2}} - i \frac{\Gamma_{3d}}{2} - I_{4p_{3/2}} - \omega}\\
\times
\left[\rho^{(3/2)}_{3/2,3/2}\frac{2}{15} + \rho^{(1/2)}_{3/2,3/2}\frac{1}{5}\right] \\
+ \frac{|\langle 4p_{3/2}^{-1}||D||3d_{3/2}^{-1}\rangle|^2}{I_{3d_{3/2}} - i \frac{\Gamma_{3d}}{2} - I_{4p_{3/2}} - \omega} 
\left[\rho^{(3/2)}_{3/2,3/2}\frac{3}{10} + \rho^{(1/2)}_{3/2,3/2}\frac{1}{30}\right]  \\
+ \frac{|\langle 4p_{1/2}^{-1}||D||3d_{3/2}^{-1}\rangle|^2}{I_{3d_{3/2}} - i \frac{\Gamma_{3d}}{2} - I_{4p_{1/2}} - \omega} 
\rho^{(1/2)}_{1/2,1/2}\frac{1}{3}  \\
+ \frac{1}{3\sqrt{10}} \langle 4p_{3/2}^{-1}||D||3d_{3/2}^{-1}\rangle \langle 4p_{1/2}^{-1}||D||3d_{3/2}^{-1}\rangle  \\
\times \biggl[\frac{\rho^{(1/2)}_{3/2,1/2}(t_{\mathrm{EUV}})}{I_{3d_{3/2}} - i \frac{\Gamma_{3d}}{2} - I_{4p_{1/2}} - \omega} \\
+ \frac{(\rho^{(1/2)}_{3/2,1/2}(t_{\mathrm{EUV}}))^{\ast}}{I_{3d_{3/2}} -
i \frac{\Gamma_{3d}}{2} - I_{4p_{3/2}} - \omega}\biggr]\Biggr\}. 
\end{multline}
Here, $\langle 4p_{j}^{-1}||D||3d_{j'}^{-1}\rangle$ is a {\em reduced} dipole matrix element \cite{Edmo57,FaRa59,Zare88}
(not to be confused with a matrix element of the reduced density matrix of the ion).  Equation (\ref{eq19}) consists of four 
distinct terms.  The first three terms are independent of the time delay (because the hole populations $\rho^{(m)}_{j,j}$ 
after the NIR pulse are constant) and describe Lorentzian line shapes associated with the three resonant transitions 
indicated in Fig.~\ref{fig1}.  The fourth term is a sum of absorptive and {\em dispersive} line shapes and depends on the 
coherence $\rho^{(1/2)}_{3/2,1/2}(t_{\mathrm{EUV}})$, which is a periodic function of the pump-probe time delay.  The period 
(6~fs \cite{SaDu06}) is defined by the energy difference between the $4p_{3/2}^{-1}$ and $4p_{1/2}^{-1}$ channels.  Note 
that for the hole dynamics to be observable in the spectrum of the transmitted radiation, it is not necessary for the relevant
resonance lines ($4p_{3/2}^{-1} \rightarrow 3d_{3/2}^{-1}$ and $4p_{1/2}^{-1} \rightarrow 3d_{3/2}^{-1}$) to spectrally 
overlap.  Only the coherent excitation of the two resonances is required, which we achieved by using an attosecond probe 
pulse, i.e., a pulse with sufficient coherent bandwidth.  

Once the EUV one-photon cross section is measured as a function of the photon energy $\omega$ and the time
delay $t_{\mathrm{EUV}}$, one can use Eq.~\eqref{eq19} to extract all nontrivial entries of the ion density matrix
[$\rho^{(3/2)}_{3/2,3/2}$, $\rho^{(1/2)}_{3/2,3/2}$, $\rho^{(1/2)}_{1/2,1/2}$, and
$\rho^{(1/2)}_{3/2,1/2}(t_{\mathrm{EUV}})$].  This is the basic idea underlying the analysis presented in
Ref.~\cite{Goulielmakis_Nature_2010}.  Such an analysis allows one to characterize (a) the degree of alignment of the
$j=3/2$ level by comparing $\rho^{(3/2)}_{3/2,3/2}$ and $\rho^{(1/2)}_{3/2,3/2}$ 
[the system is fully aligned if $\rho^{(3/2)}_{3/2,3/2}=0$ and $\rho^{(1/2)}_{3/2,3/2} \ne 0$, and is unaligned if 
$\rho^{(3/2)}_{3/2,3/2} = \rho^{(1/2)}_{3/2,3/2}$]; (b) the population of the $j=1/2$ level
relative to the population of the $j=3/2$ level [$\rho^{(1/2)}_{1/2,1/2}/(\rho^{(3/2)}_{3/2,3/2} +
\rho^{(1/2)}_{3/2,3/2})$]; and (c) the degree of coherence by calculating
$|\rho^{(1/2)}_{3/2,1/2}|/\sqrt{\rho^{(1/2)}_{3/2,3/2}\rho^{(1/2)}_{1/2,1/2}}$.

In order to be able to determine the ion density matrix elements using Eq.~\eqref{eq19}, the reduced dipole matrix elements 
must be known.  To this end, we proceeded as follows.  Using GRASP \cite{PaFr96}, we calculated the oscillator strengths
for transitions from the $3d_{j}^{-1}$ levels to the $4p_{j'}^{-1}$ levels.  In this way, we obtained
$|\langle 4p_{3/2}^{-1}||D||3d_{5/2}^{-1}\rangle|^2=0.119$~a.u.,
$|\langle 4p_{3/2}^{-1}||D||3d_{3/2}^{-1}\rangle|^2=0.0126$~a.u., and 
$|\langle 4p_{1/2}^{-1}||D||3d_{3/2}^{-1}\rangle|^2=0.0695$~a.u.  The relative ratios are very close to what would be 
obtained within the LS coupling scheme.  This allowed us to employ standard angular-momentum algebra within the LS coupling
scheme \cite{Edmo57,FaRa59,Zare88} to determine the relative sign between $\langle 4p_{3/2}^{-1}||D||3d_{3/2}^{-1}\rangle$
and $\langle 4p_{1/2}^{-1}||D||3d_{3/2}^{-1}\rangle$.  Thus, $\langle 4p_{3/2}^{-1}||D||3d_{3/2}^{-1}\rangle 
\langle 4p_{1/2}^{-1}||D||3d_{3/2}^{-1}\rangle = -0.0296$.  The natural lifetime broadening from the Auger decay of the
$3d_{j}^{-1}$ levels is $\Gamma_{3d} = 0.00323$~a.u. \cite{JuKi01}.  We mention that, when analyzing experimental transient 
absorption data, the spectrometer resolution must also be taken into consideration.  

\begin{table}
  \caption{\label{tab1} Normalized density matrix elements of strong-field-generated Kr$^+$ ions (a) extracted from 
    experiment \cite{Goulielmakis_Nature_2010} using Eq.~\eqref{eq19} and (b) calculated using the theory from Ref.~\cite{RoSa09} 
    assuming the NIR pulse parameters specified in Ref.~\cite{Goulielmakis_Nature_2010}.}
\begin{ruledtabular}
\begin{tabular}{c|c|c}
& (a) experiment & (b) theory \\
\hline
$2\rho^{(3/2)}_{3/2,3/2}$ & 0.23 & 0.05 \\
$2\rho^{(1/2)}_{3/2,3/2}$ & 0.42 & 0.69 \\
$2\rho^{(1/2)}_{1/2,1/2}$ & 0.35 & 0.26 \\
$|\rho^{(1/2)}_{3/2,1/2}|$ & 0.12 & 0.13
\end{tabular}
\end{ruledtabular}
\end{table}

Based on the approach just described, a fit was performed in Ref.~\cite{Goulielmakis_Nature_2010} to determine, directly from
the experimental attosecond transient absorption data, the reduced density matrix elements of strong-field-generated krypton 
ions.  The results are collected in Table~\ref{tab1}.  Also shown in Table~\ref{tab1} are the ion density
matrix elements calculated using the time-dependent multichannel theory developed in Ref.~\cite{RoSa09}.  The NIR pulse
parameters assumed in the calculation were taken from Ref.~\cite{Goulielmakis_Nature_2010}.  Note that the ion density
matrix elements have been normalized such that the trace of the ion density matrix equals unity.  It is evident from
Table~\ref{tab1} that experiment and theory give a similar degree of coherence.  Also, experiment and theory give a
similar value for the population of the $j=1/2$ level relative to the population of the $j=3/2$ level.  However, the
experimental data suggest a much smaller degree of alignment of the $j=3/2$ level than predicted by our theory.

\begin{figure}[ht]
  \begin{center}
    \includegraphics[clip,width=0.8\hsize,angle=0]{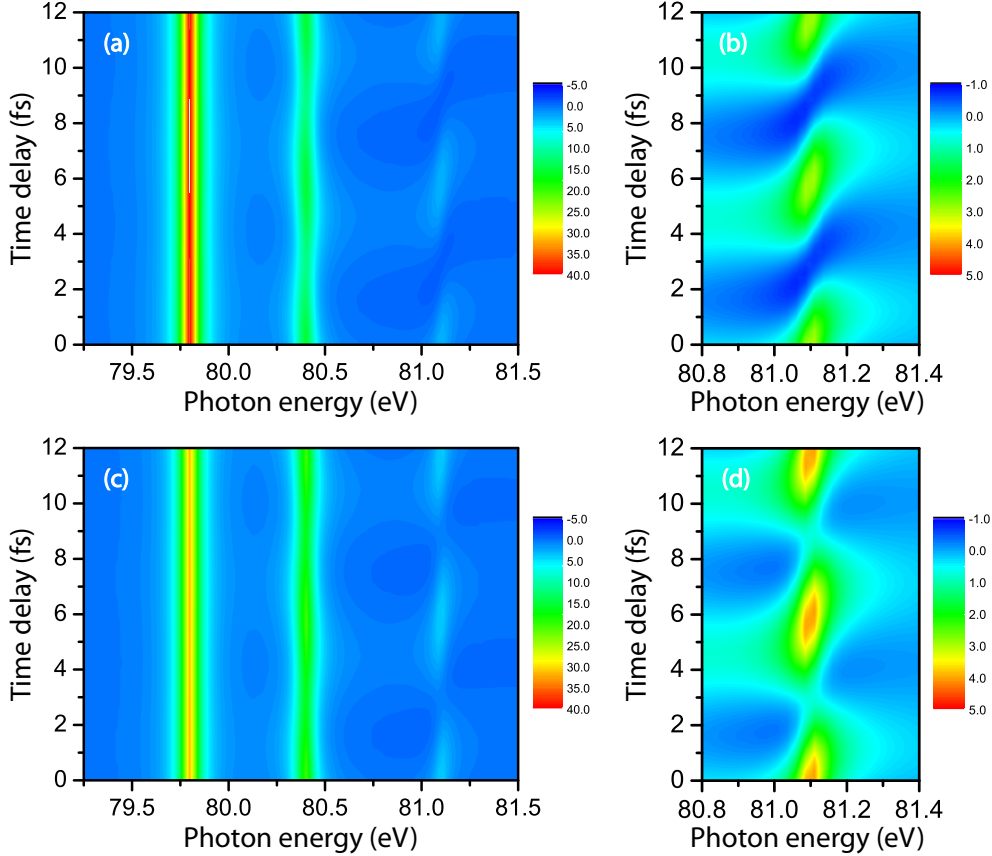}
    \caption{(Color online) Attosecond transient absorption cross section, in Mb, of strong-field-generated Kr$^+$,
      plotted as a function of the photon energy and the time delay.
      The cross section was calculated using Eq.~\eqref{eq19}.
      Panels (a) and (b) show the attosecond transient absorption cross section for the 
      theoretical ion density matrix elements in Table~\ref{tab1}.  In panels (c) and (d),
      the attosecond transient absorption cross section is shown for the experimental ion density 
      matrix elements in Table~\ref{tab1}.}
      \label{fig2}
  \end{center}
\end{figure}

The attosecond transient absorption cross sections [Eq.~\eqref{eq19}] for the two sets of density matrix elements are
plotted in Fig.~\ref{fig2}. Again, there is no overlap between the NIR and EUV pulses, that is, the NIR pulse is
  centered at a large negative value of $t_\mathrm{EUV}$.  The strongest of the three absorption lines does not depend on
the time delay.  This absorption line corresponds to a transition to the $3d_{5/2}^{-1}$ level.  Since this level can be
reached only from the $4p_{3/2}^{-1}$ level, but not from the $4p_{1/2}^{-1}$ level (see Fig.~\ref{fig1}), the strongest
absorption line is insensitive to the coherence between $4p_{3/2}^{-1}$ and $4p_{1/2}^{-1}$.  The other two transitions,
which both involve the $3d_{3/2}^{-1}$ level (see Fig.~\ref{fig1}), display a conspicuous dynamical behavior as a function
of the time delay. The modulation of the weakest line, which corresponds to the transition $4p_{3/2}^{-1} \rightarrow
3d_{3/2}^{-1}$, is most pronounced due to the coupling to the relatively strong $4p_{1/2}^{-1} \rightarrow 3d_{3/2}^{-1}$
transition. Moreover, for certain delays the absorption cross section for that line even becomes negative. This happens
when the coherent population transfer from $4p_{1/2}^{-1}$ to $4p_{3/2}^{-1}$ via $3d_{3/2}^{-1}$ dominates over the
absorption from the $4p_{3/2}^{-1}$ state.

Apart from the oscillation of the respective line strengths
% Not only do their respective strengths oscillate 
with the spin-orbit period of 6 fs, the energetic positions of the resonance lines oscillate as well.  
This is particularly easy to see in panels (b) and (d) in Fig.~\ref{fig2}.  These
energy oscillations are a consequence of the interplay between the absorptive and dispersive terms mentioned earlier
in connection with Eq.~\eqref{eq19}.

\section{The accuracy of Beer's law}
\label{sec4}

In the previous section, our discussion was based on the approximation that different frequency components of the EUV
pulse propagate independently. The assumption $\tilde{P}^{(1)}(x,\omega) \propto
\tilde{\mathcal{E}}_{\mathrm{EUV}}(x,\omega)$ allowed us to describe the polarization response with a linear
susceptibility $\chi^{(1)}(\omega)$ and to integrate Eq.~\eqref{eq13f} analytically. In this section, we investigate
the accuracy of this approximation.

As follows from Eq.~\eqref{eq:P2},
% As it was mentioned in section~\ref{sec2b},
the polarization response $\tilde{P}^{(1)}(x,\omega)$ is not proportional to $\tilde{\mathcal{E}}_{\mathrm{EUV}}(x,\omega)$
if the density matrix contains non-zero off-diagonal elements that correspond
to states coupled by dipole transitions through an intermediate excited state.  Physically, this means that an ion that
absorbs a photon with an energy $\omega_1$ can \emph{coherently} emit a photon with a different energy $\omega_2$,
provided that the initial ionic state is a \emph{coherent} superposition of two or more states. In this typical
$\Lambda$-scheme, the polarization response at the photon energy $\omega_2$ obviously depends not only on
$\tilde{\mathcal{E}}_{\mathrm{EUV}}(x,\omega_2)$, but also on $\tilde{\mathcal{E}}_{\mathrm{EUV}}(x,\omega_1)$. In this
case, Eq.~\eqref{eq13g} is an approximation.

To go beyond this approximation, we numerically solve the first-order propagation equation \eqref{eq13f}
using Eq.~\eqref{eq:P2} to evaluate the polarization response of the medium at each propagation step.  Even though 
$\tilde{P}^{(1)}(x,\omega)$ is no longer proportional to $\tilde{\mathcal{E}}_{\mathrm{EUV}}(x,\omega)$, we compare the
results of numerical propagation with those obtained in the previous section
in terms of the \emph{apparent} one-photon cross section:
\begin{equation}
\label{eq:apparent_sigma}
\sigma_\mathrm{app}(\omega) = \frac{1}{n_{\mathrm{AT}} L}
\ln \frac{\left|
    \tilde{\mathcal{E}}_{\mathrm{EUV}}(0,\omega) \right|^2}
{\left| \tilde{\mathcal{E}}_{\mathrm{EUV}}(L,\omega) \right|^2}.
\end{equation}

\begin{figure}[ht]
  \begin{center}
    \includegraphics[clip,width=0.8\hsize,angle=0]{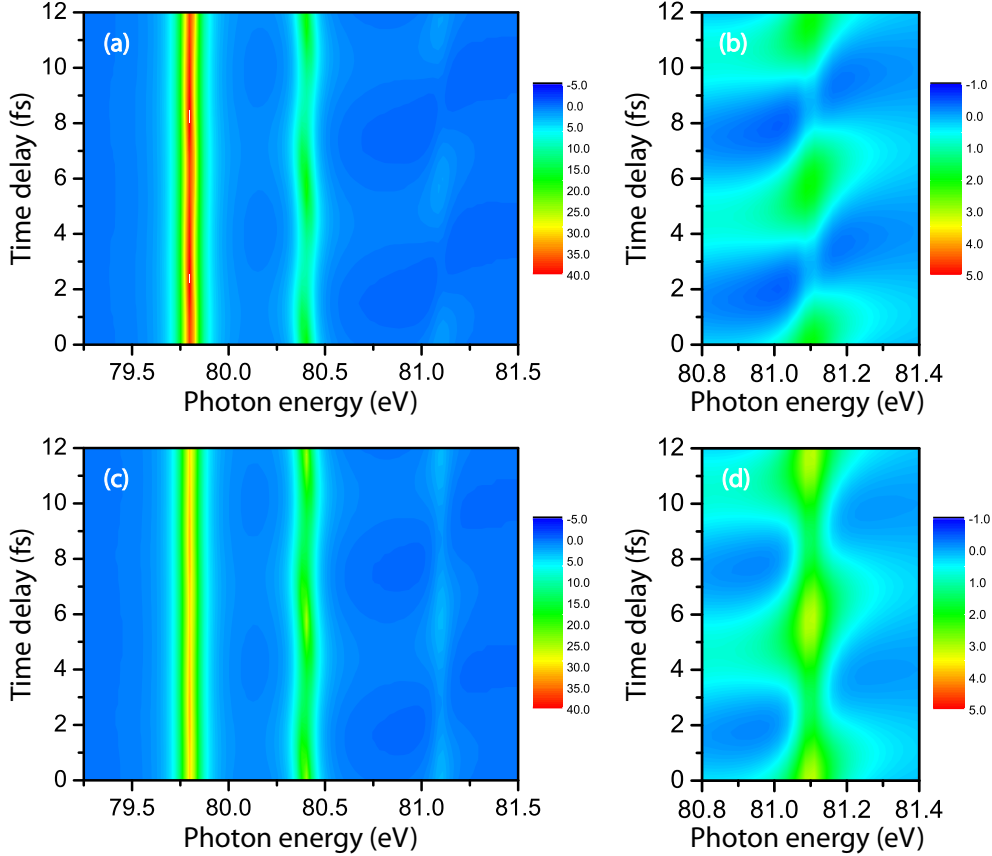}
    \caption{(Color online) Apparent attosecond transient absorption cross section, in Mb, of strong-field-generated 
      Kr$^+$, plotted as a function of the photon energy and the time delay.
      The cross section was calculated by applying Beer's law to the numerically propagated EUV field.
      Panels (a) and (b) show the apparent attosecond transient absorption cross section for the
      theoretical ion density matrix elements in Table~\ref{tab1}.  In panels (c) and (d),
      the apparent attosecond transient absorption cross section is shown for the experimental ion density
      matrix elements in Table~\ref{tab1}.}
    \label{fig3}
  \end{center}
\end{figure}

\begin{figure}[ht]
  \begin{center}
    \includegraphics[clip,width=0.8\hsize,angle=0]{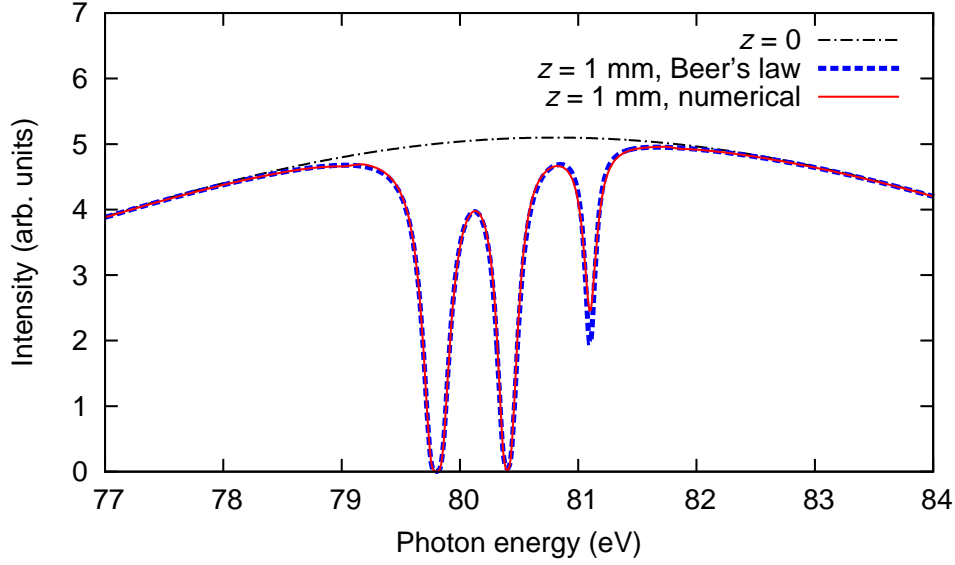}
    \caption{(Color online) The spectrum of the EUV pulse before (dash-dotted line) and after propagation for $t_\mathrm{EUV}=0$. 
      The dashed line, representing a simulation assuming the validity of Beer's law, corresponds to panels (c) and (d) in
      Fig.~\ref{fig2}. The solid line shows the result of numerical propagation, corresponding to panels (c) and (d) in
      Fig.~\ref{fig3}.}
    \label{fig4}
  \end{center}
\end{figure}

In Fig.~\ref{fig3}, we show a false-color representation of $\sigma_\mathrm{app}(\omega)$ evaluated for different
delays between the NIR pump and EUV probe pulses.
For this simulation, we used a bandwidth-limited Gaussian EUV pulse with a central photon energy of 80.8~eV and a full
width at half maximum of intensity equal to 150 attoseconds. The elements of the density matrix were taken from
Table~\ref{tab1}. The propagation in a gas of strong-field-generated $\mbox{Kr}^+$ ions with an atomic number density 
of $n_{\mathrm{AT}}=2.2 \times 10^{18}\ \mbox{cm}^{-3}$ was terminated after $L=1$~mm.
% For comparison, the absorption length $L_\mathrm{abs}(\omega) = 1/[n_\mathrm{AT} \sigma^{(1)}(\omega)]$ at $\hbar\omega = 79.8$~eV, 80.4~eV, and 81.1~eV is equal to ???, ???, and ???, respectively.
The spectrum of the EUV pulse before and after propagation, for the experimental ion density
matrix elements in Table~\ref{tab1}, is shown in Fig.~\ref{fig4}.

A careful inspection of Figs.~\ref{fig2}, \ref{fig3}, and \ref{fig4} reveals that the approximation underlying 
Beer's law notably affects the line at 81.1~eV, which corresponds to the transition $4p_{3/2}^{-1}
\rightarrow 3d_{3/2}^{-1}$. This is not surprising: for the same reasons why this absorption line exhibits strong quantum
beats, it is also sensitive to other effects related to off-diagonal elements of the density matrix. Still, in spite of
the strong absorption, the discrepancy between the model assuming the validity of Beer's law and the results obtained by
numerically solving the propagation equation is rather small. 

It is instructive to repeat the analysis based on
Eq.~\eqref{eq19} and retrieve the density matrix from the apparent absorption cross section
as if Beer's law were rigorously valid.
Table~\ref{tab2} gives these apparent density matrix elements. The good agreement between
Tables~\ref{tab1} and \ref{tab2} indicates that Beer's law is indeed a good approximation for extracting
electronic-structure information from transient absorption data, even though it should be used with care.
We may conclude, in particular, that the discrepancy between the experimental and theoretical degrees of
ion alignment ($j=3/2$ level) cannot be explained by a failure of Beer's law.

\begin{table}
\caption{\label{tab2} Apparent density matrix elements of strong-field-generated Kr$^+$ ions extracted from
the apparent attosecond transient absorption cross section for (a) the experimental ion density matrix elements 
in Table~\ref{tab1} and (b) the theoretical ion density matrix elements in Table~\ref{tab1}.}
\begin{ruledtabular}
\begin{tabular}{c|c|c}
& (a) & (b) \\
\hline
$2\rho^{(3/2)}_{3/2,3/2}$ & 0.21 & 0.03 \\
$2\rho^{(1/2)}_{3/2,3/2}$ & 0.44 & 0.71 \\
$2\rho^{(1/2)}_{1/2,1/2}$ & 0.35 & 0.26 \\
$|\rho^{(1/2)}_{3/2,1/2}|$ & 0.14 & 0.14
\end{tabular}
\end{ruledtabular}
\end{table}

\section{Conclusions}
\label{sec5}

In this paper, we discussed the theory underlying attosecond transient absorption spectroscopy of strong-field-generated
ions.  This theory was employed in Ref.~\cite{Goulielmakis_Nature_2010} to experimentally determine the 
reduced density matrix of Kr$^+$ ions produced by an intense NIR pulse.  Good agreement between experiment and theory
was found for the degree of coherence and for the population of the $j=1/2$ level relative to the population of the 
$j=3/2$ level.  However, experiment suggests strongly suppressed alignment of the $j=3/2$ level, which is not 
consistent with calculations based on the theory described in Ref. \cite{RoSa09}.  The origin of this discrepancy is 
currently unknown.  

As demonstrated in this paper, EUV propagation effects beyond Beer's law do not explain the discrepancy between 
experiment and theory found for few-cycle NIR pulses \cite{Goulielmakis_Nature_2010}.  Earlier measurements on Kr$^+$ 
ions generated using 50 fs NIR pulses gave a degree of alignment in rather good agreement with an adiabatic 
strong-field-ionization theory \cite{SoAr07}.  A noticeable reduction of ion alignment, in comparison to the adiabatic
strong-field-ionization theory, was observed in Xe$^+$ ions generated using 45 fs NIR pulses \cite{LoKh07}.
But the effect was not as pronounced as it is in the current case, and it was surmised at the time that the discrepancy
is a consequence of nonadiabatic effects \cite{LoKh07}.  However, nonadiabatic effects cannot explain the disagreement
between the experimental and theoretical ion density matrix elements shown in Table~\ref{tab1}, for the 
strong-field-ionization theory employed \cite{RoSa09} is based on numerical wave-packet propagation and 
does not suffer from the limitations of the adiabatic approximation.  

It seems likely that the origin of the discrepancy must be sought in either of the following two possibilities.
The first possibility is that in the experiment of Ref. \cite{Goulielmakis_Nature_2010}, multielectron effects 
beyond the multichannel theory of Ref. \cite{RoSa09} played an important role.  Since, in view of Ref. \cite{SoAr07}, 
these multielectron effects would appear to have a smaller impact when using longer NIR pulses, the observed discrepancy 
might suggest an enhancement of multielectron effects by few-cycle pulses.  The second possibility is that 
the experiment of Ref. \cite{Goulielmakis_Nature_2010} was affected by substantial NIR propagation effects.  This could be
clarified by repeating the experiment at a lower target density.

\acknowledgments
We thank E. Goulielmakis, N. Rohringer, D. Charalambidis, S. R. Leone, and F. Krausz for fruitful discussions.
This research was supported in part by the National Science Foundation under Grant No. NSF PHY05-51164.
V.~Y. acknowledges support by the DFG Cluster of Excellence: Munich-Centre for
Advanced Photonics (MAP). T.~P. acknowledges support by an MPRG grant of the Max-Planck-Gesellschaft.
Z.-H.~L. is supported by the National Science Foundation (CHE-0742662 and EEC-0310717).

\end{document}